\documentclass{article}

\usepackage{arxiv}

\usepackage{graphicx} 
\usepackage{todonotes}
\usepackage{geometry}
\usepackage{pdflscape}
\usepackage{longtable}
\usepackage{booktabs}
\usepackage{array}
\usepackage{float}
\usepackage{subcaption}
\usepackage[export]{adjustbox}
\usepackage{wrapfig}
\usepackage{placeins}
\usepackage[backend=biber,style=authoryear]{biblatex}
\title{It's About Time: The Copilot Usage Report 2025 \\ 
\vspace{0.5em} 
\normalsize The Temporal and Modal Dynamics of Copilot Usage}

\author{%
\begin{tabular}{c}
\textbf{Beatriz Costa-Gomes} \quad \textbf{Sophia Chen} \quad \textbf{Connie Hsueh} \quad \textbf{Deborah Morgan} \\
\textbf{Philipp Schoenegger} \quad  \textbf{Yash Shah} \quad \textbf{Sam Way} \quad \textbf{Yuki Zhu}\\
\textbf{Timoth\'e Adeline} \quad \textbf{Michael Bhaskar} \quad \textbf{Mustafa Suleyman} \quad \textbf{Seth Spielman}\\[0.6em]
Microsoft AI
\end{tabular}
}

\date{December 2025}

\begin{document}

\maketitle

\begin{abstract}
We analyze 37.5 million deidentified conversations with Microsoft's Copilot between January and September 2025. Unlike prior analyses of AI usage, we focus not just on what people do with AI, but on how and when they do it. We find that how people use AI depends fundamentally on context and device type. On mobile, health is the dominant topic, which is consistent across every hour and every month we observed—with users seeking not just information but also advice. On desktop, the pattern is strikingly different: work and technology dominate during business hours, with ``Work and Career'' overtaking ``Technology'' as the top topic precisely between 8 a.m. and 5 p.m. These differences extend to temporal rhythms: programming queries spike on weekdays while gaming rises on weekends, philosophical questions climb during late-night hours, and relationship conversations surge on Valentine's Day. These patterns suggest that users have rapidly integrated AI into the full texture of their lives, as a work aid at their desks and a companion on their phones.
\end{abstract}

\section{Introduction} 

AI models are increasingly embedded in many aspects of modern work and life, with at least 10\% of the adult population engaging with AI on a weekly basis \parencite{chatterji2025people}. While much of the early work on AI usage emphasized the use of AI for productivity related tasks \parencite{brynjolfsson2025generative}, recent work has shown a rise in non-work related tasks. While we have an understanding of ``what'' people do with AI, we know less about when and how they do it. Analyzing how and when people use Copilot is critical for understanding the current economic and social impacts of AI, and what the future trajectory of AI adoption may look like.  

Generative AI has diffused more rapidly than any other technology in history \parencite{bick2024rapid, aiei_aidiffusion_2025}.  However, diffusion unfolds at multiple scales, spreading both through the economy and people's daily lives \parencite{rogers2014diffusion, wejnert2002integrating,hagerstrand1968innovation}.  Only by understanding how people relate to these technologies can we truly build  humanistic AI.  In this large-scale analysis of Copilot usage we focus not just on what people do with AI, but on how and when they do it.

Recent work drawing on large-scale chat data has primarily focused on work use cases. For example, OpenAI \parencite{chatterji2025people} analyzed 1.1M sampled ChatGPT messages (May 2024--July 2025) with privacy-preserving, prompt-defined classifiers to label work status, intent, topic, and O*NET (a database of occupations) activity. They documented a rise in non-work messages (53\%\,$\to$\,73\% year over year), a dominance of practical guidance, seeking information, and writing (about 77\% of conversations, with writing at about 40\% of work use in June 2025), and an intent mix led by asking (51.6\%) over doing (34.6\%) and expressing (13.8\%). Similarly focusing on work, Anthropic \parencite{handa2025economic} drew on over 4M Claude.ai conversations (December 2024--January 2025) to look at AI usage at work. They found that usage of Claude is concentrated in IT/technology (\(\sim\)50\%), followed by creative/cultural (\(\sim\)20\%) and business/finance/customer service (\(\sim\)15\%) use cases. In general, adoption is broad but shallow across occupations, and augmentative modes with necessary human guidance are more common than automation (57\% vs.\ 43\%). Automation is concentrated in content generation and coding ``fix-and-retry'', while augmentation is more common in front-end development, professional communication, education, and validation. 

Previous work from Microsoft analyzing 200,000 conversations from an earlier iteration of Copilot by Bing finds that the most common work activities that AI assists with are gathering information and writing \parencite{tomlinson2025working}. Further work has also documented domain-specific AI use among students, programmers, and clinicians \parencite{ammari2025students,klemmer2024using,stackoverflow_dev_survey_2025_ai,mandal2025utilization}, with specific use patterns emerging for different groups. Looking at online forum data, \textcite{zao-sanders2025-hbr-genai-usage} reported a shift toward emotional/support use cases, with technical troubleshooting declining.

Despite this progress in understanding AI usage in the context of work tasks and productivity, two basic dimensions of use remain underexplored: how modality (mobile vs.\ desktop) and time dynamics shape usage. There is already a rich literature on how technology relates to and is integrated into daily life in general \parencite{silverstone1992moral}. Personal computing and mobile phones are notable instances of technology development and integration of the last few decades, and each iteration has impacted how and when users engage, access and shape information through their devices \parencite{kamvar2009computers, booth2023mental, garg2020he}. Prior work has not fully analyzed how AI use differs by device type or how patterns evolve temporally across the day or year.

We do not claim that the complex relationship between people and AI can be fully revealed by a classification of conversation logs.  However, we do think that taking a look at how and when people use AI can provide insight into the social integration of the technology \parencite{bohmer2011falling}. We extend prior work by teasing out differences in AI usage by device type and time. We believe that this disambiguation enables a deeper understanding of people's relationship with AI. 

This paper is based on 37.5M deidentified conversations randomly sampled on a daily basis between January and September 2025 from Copilot. This allows us to analyze how AI usage has evolved over both the course of the year and within each day, across both mobile and desktop devices. These patterns illustrate that Copilot is utilized as a productivity tool when users are sitting at their work desktops, yet shifts into a conversational partner on questions of philosophy, health, and wellness when users are on their phones at night.

\section{Classifying Copilot Usage}
In line with industry best practice, Microsoft analyzes Copilot usage metrics in a privacy-preserving way, by running machine-based classifiers over deidentified conversation logs that are automatically scrubbed of any personally identifiable information (PII) \parencite{tamkin2024clioprivacypreservinginsightsrealworld, chatterji2025people}. All data used in this paper was processed in accordance with Microsoft's privacy policy \parencite{MicrosoftPrivacy2025}. Our machine-based classifiers are ``eyes-off'', i.e., no human researcher sees the content of a conversation.

The data used in this analysis is from a random sample of user interactions with Copilot from January 7 to September 25, 2025. We exclude all enterprise traffic and users that have signed in on a commercial or educational enterprise account. The number of daily conversation samples was generally consistent throughout the collection period. We sample around 144,000 conversations per day from January 7 to September 25, 2025 ($\approx$ 260 days). Conversations are sampled randomly and thus are representative of Copilot usage. Sampled conversations include coarse location (country), date, time, and the application on which the interaction occurred. Information about the application being used allows us to classify the usage as taking place either on a mobile device or a personal computer/desktop. Each sampled conversation is labeled by machine-based classifiers that assign the conversation a \textit{topic} and an \textit{intent} tag, or in the cases of ambiguous messages and low classifier certainty a no-topic/no-intent tag.  These ``No Topic''/``No Intent'' conversations are mostly single or short conversations for which we cannot determine the user's goal or subject of conversations (e.g. ``Hi Copilot'') at sufficient certainty. These kinds of vague, short, exploratory conversations were common early in the year but diminished significantly over the study period. 

Topics are the subjects of conversation. While there are many sets of possible topics, our list of topics adapts widely used Microsoft systems and third-party taxonomies for classifying web search activities to conversational AI \parencite{topics_api_2024} (see Table \ref{tab:topics}). 

Intents are the ``jobs to be done'', or the tasks that the user would like to accomplish with Copilot. The intents for Copilot were developed via user research to capture tasks that are common on conversational AI systems. For our list of intents, see Table \ref{tab:intent}.

Each conversation is assigned a topic and intent independently. In total, this yields a system of $\sim$300 topic-intent pairs. For example, ``Seeking Advice'' is an intent, while ``Relationships'' and ``Work and Career'' are topics. A user can be seeking advice about relationships or seeking advice about work and career, which would be captured by these respective topic/intent pairs. While far from exhaustive, the combination of topics and intents provides an overview of Copilot usage.

\begin{center}
\begin{longtable}{>{\raggedright\arraybackslash}p{5cm} >{\raggedright\arraybackslash}p{8cm}}
\caption{Conversation topics and descriptions} \label{tab:topics} \\
\toprule
\textbf{Topic} & \textbf{Description (what the conversations are about)} \\
\midrule
\endfirsthead
\toprule
\textbf{Topic} & \textbf{Description (what the conversations are about)} \\
\midrule
\endhead
Art and Design & Art, design, and creative projects. \\
Autos and Vehicles & Cars, motorcycles, and other vehicles. \\
Beauty and Fashion & Beauty and fashion. \\
Books and Literature & Books, literature, and reading. \\
Business and Industry & Business or industry such as strategy, planning, and entrepreneurship. \\
Education & Education, learning, and school. Excludes conversations about specific school subjects that have their respective topics. \\
Entertainment & Movies, music, TV shows, and other forms of entertainment. Excludes conversations about video games, and other forms of games which should be in Games. \\
Food and Drink & Food, cooking, and beverages. \\
Games & Video games, role-playing games, and other forms of gaming. \\
Health and Fitness & Health, fitness, and medical concerns. \\
Hobbies and Leisure & Hobbies, crafts, and leisure activities. Excludes conversations about Games, Books and Literature, Entertainment, Travel, Sports and other hobbies/leisure activities that have an assigned topic. \\
Home and Garden & Home improvement, gardening, and DIY projects. \\
Internet and Social Media & Internet, social media. Excludes conversations covered in other topics, like News and Current Events. \\
Language Learning and Translation & Learning a natural language or simple translation requests. Excludes conversations about programming languages, which should be in Programming. \\
Local & Nearby events, restaurants, institutions, attractions.  \\
Math and Logic & Mathematics and logic. \\
Money & Money, personal finance, and investments. \\
News and Current Events & News, current events of regional, national, or global interest, and politics. \\
Personal Growth and Wellness & Mental health, personal development, self-improvement, and self-help. \\
Pets and Animals & Pets, animals, and wildlife. \\
Programming & Computer programming. \\
Relationships & Relationships, family, dating, and love. \\
Religion and Philosophy & Religion, spirituality, and philosophy. \\
Science & Science and scientific research. \\
Society, Culture, and History & Society, culture, and history. Excludes conversations about news and current events. \\
Sports & Professional sports. \\
Technology & Technology, gadgets, and software. \\
Travel & Travel, vacations, and destinations. \\
Work and Career & Work, career, and professional development. \\
No Topic & A ``No Topic'' conversation has a vague or unclear subject. \\
\bottomrule
\end{longtable}
\end{center}

\begin{center}
\begin{longtable}{>{\raggedright\arraybackslash}p{4cm} >{\raggedright\arraybackslash}p{8cm}}
\caption{Conversation intents and descriptions. Bold corresponds to the name used for the intent.} \label{tab:intent} \\
\toprule
\textbf{Intent} & \textbf{Description} \\
\midrule
\endfirsthead
\toprule
\textbf{Intent} & \textbf{Description} \\
\midrule
\endhead
\textbf{Creating} Text or Images & Creating text, imagery, code, or any content from scratch. \\
\textbf{Editing} Text or Images & Editing text, imagery, code, or any content from an original source. \\
\textbf{Entertainment} & Using the AI to pass-time, play games, have undirected chit-chat types of conversations. In Entertainment conversations, often the conversation itself is the goal. \\
Getting Feedback or \textbf{Advice} & Getting feedback, advice, emotional support, or brainstorming on anything NOT related to Shopping and Product Research. \\
Information \textbf{Summary} & Asking AI to summarize content, extract key points. If the summarization task is to help a user learn a new skill or related to schoolwork, it should be in the Learning category; if it is about summarizing products, it should be in the Shopping and Product Research category. \\
\textbf{Learning} & Helping the user with schoolwork or learning a new skill. \\
\textbf{Planning} & Help with planning tasks, travel, or events; organizing one's time for increased efficiency. \\
\textbf{Searching} for Information & Searching for general information about a topic, business, person, and/or place. Seeking information related to products or services that one might purchase, including product comparisons and reviews, should be in the Shopping and Product Research category. Questions related to schoolwork or learning a new skill should be in the Learning category. \\
\textbf{Shopping} and Product Research & Searching for information about a product or service. Assisting with shopping decisions, especially learning about products, categories of products, or comparing products. \\
\textbf{Technical Support} & Help with computers, electronics, and other devices. Providing technical support for software, hardware, or other technical issues. \\
No Intent & A ``No Intent'' conversation is a short exchange between the user and the AI for which we cannot determine the user's goal(s). \\
\bottomrule
\end{longtable}
\end{center}

\section{How and When People Use Copilot}
\subsection{The General Overview}

We look at Copilot usage by ranking topics and intents: the most common topic of conversation for that time is marked as rank 1, and the least common as rank 30. Intents are ranked 1--11. For topic-intent pairs, we consider the rank of the conversation for each pair, resulting in ranks ranging from 1 to 330. 

In Table \ref{tab:ranks}, we show the top 5 topics and top 5 intents across the entire dataset. We find that the most common topics of conversation on Copilot are ``Technology'' as well as ``Work and Career'', suggesting strong productivity-aimed uses of Copilot. However, conversations about ``Health and Fitness'' also rank highly, indicating that people also turn to Copilot for more personal questions. ``Searching'' is the most common intent across conversations, followed by ``Advice''.

\begin{table}
\caption{General top 5 ranks for Topic and Intent.} \label{tab:ranks}
\centering
\begin{tabular}{|| c | c ||}
 \hline
 Rank & Topic \\ [0.5ex] 
\hline \hline
1 & Technology\\
2 & Work and Career	\\
3 & Health and Fitness\\
4 & Language learning and translation\\
5 & Society, Culture, and History \\
[1ex] 
 \hline
\end{tabular}
\quad
\begin{tabular}{|| c | c ||}
 \hline
 Rank & Intent \\ [0.5ex] 
\hline \hline
1 & Searching \\
2 & Advice  \\
3 & Creating \\
4 & Learning \\
5 & Technical Support \\
[1ex] 
 \hline
\end{tabular}
\end{table}

\begin{figure}[p]
  \centering
  \includegraphics[width=\linewidth]{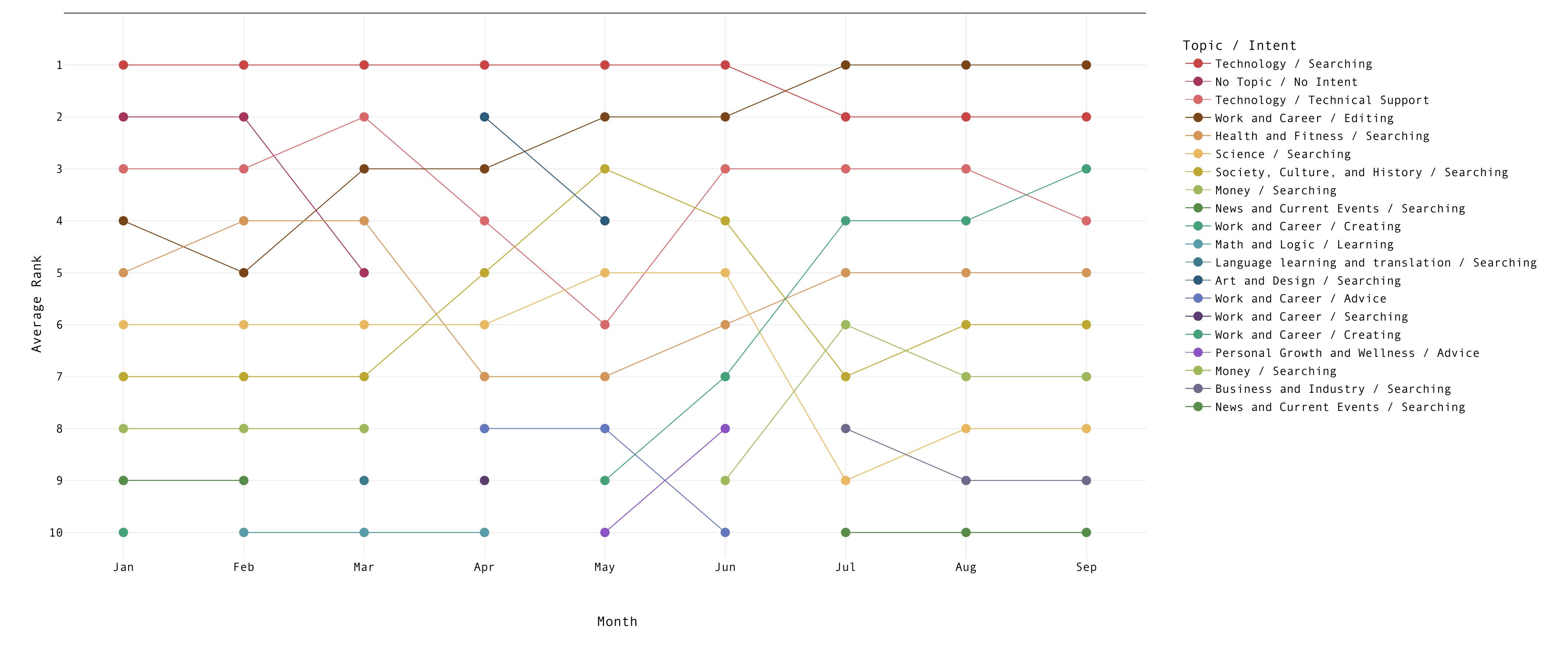}
  \caption{Monthly top 10 ranks for topic-intent pairings on desktop. The lines correspond to when the pairing is present in consecutive  months.}
  \label{fig:desktop_tip}
\end{figure}

\begin{figure}[p]
  \centering
  \includegraphics[width=\linewidth]{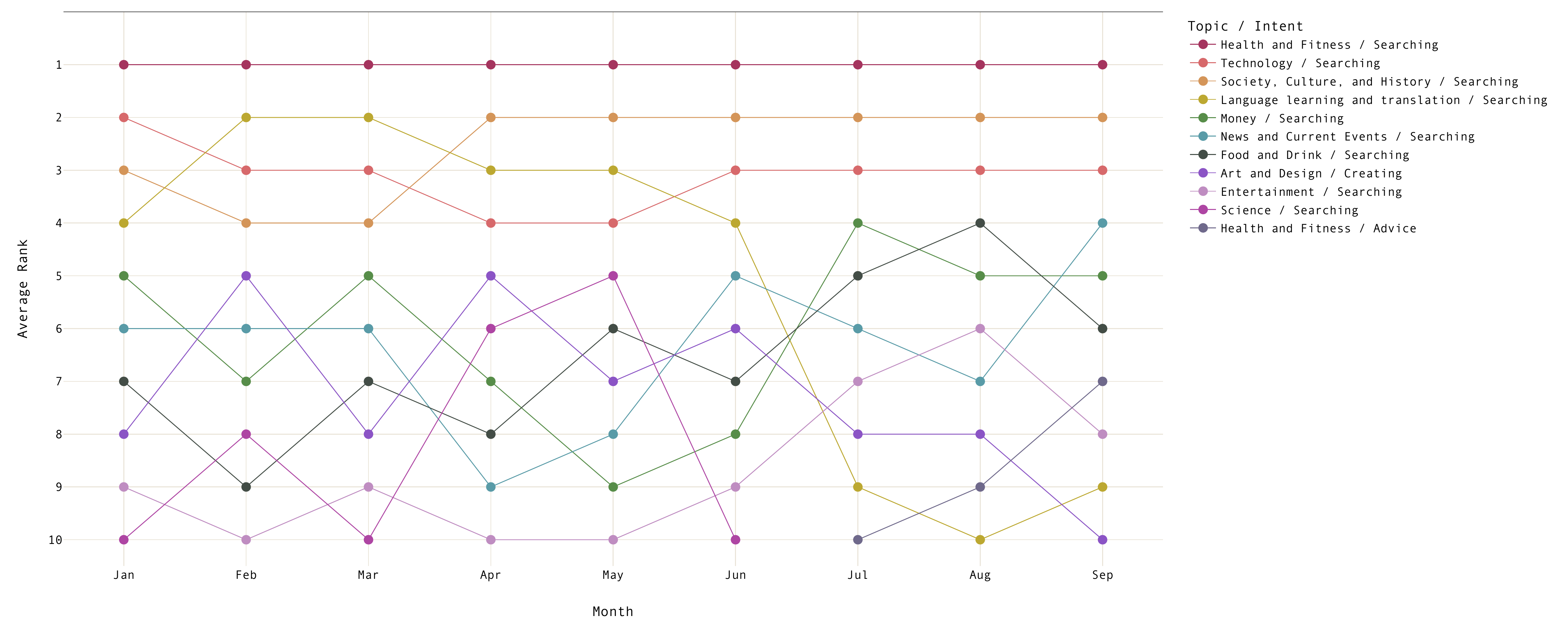}
  \caption{Monthly top 10 ranks for topic-intent pairings on mobile. The lines correspond to when the pairing is present in consecutive months.}
  \label{fig:mobile_tip}
\end{figure}

We find substantial differences between how AI is used on different types of devices, and there has been more volatility in usage patterns on desktops and laptops than on mobile devices, as seen in Figures \ref{fig:desktop_tip} and \ref{fig:mobile_tip}. The most frequent topic-intent pairs appear more stable on mobile; ``Health and Fitness'' / ``Searching'' is unchanged as the most frequent pair, and four pairs remain within the top four for the majority of the year. While the stability in usage is an interesting observation, we think that dominance of ``Health and Fitness'' on mobile, regardless of time of day or date suggests that users engage with Copilot as a confidant for personal topics and as a companion for personal improvement. This rise of advice‑seeking conversations further highlights a growing user trust in Copilot, as individuals increasingly view it not only as a source of information but as a reliable source of advice.  On personal computers, Copilot usage has been more volatile than on phones; 20 distinct topic-intent pairs entered the top 10 rank over the course of the year, showing that relative use of Copilot is changing month-to-month at a high rate. This is compared to 11 distinct topic-intent pairs for Copilot mobile use over the year. 

On desktop, we can see how several intents are paired with the ``Work and Career'' topic, which does not feature at all on mobile. Given that our dataset explicitly excludes enterprise-authenticated traffic, this finding highlights the permeability of the boundary between professional and personal computing. Interestingly, ``Art and Design / Creating'' holds ranks 3 and 4 on desktop for two months, and then disappears from the top 10 while it consistently remains in the top 10 on mobile, with a drop only in September.  This pattern might be related to popular public interest around image generation at this time in 2025.  In the summer months, we observe certain subjects related to school work declining in prevalence, such as ``Language learning and Translation'' and ``Science'', with ``Entertainment'' increasing in rank. 


Overall, the data suggests that, as might be expected, Copilot users reach for their phones for the topics that are more personal or pique their interest beyond work. The consistency is nonetheless marked. However, to provide a complete picture, we must also examine the temporal dynamics of usage. 

\subsection{Temporal Dynamics of Copilot Usage}

\begin{figure}[htbp]
  \centering
  \includegraphics[width=\linewidth]{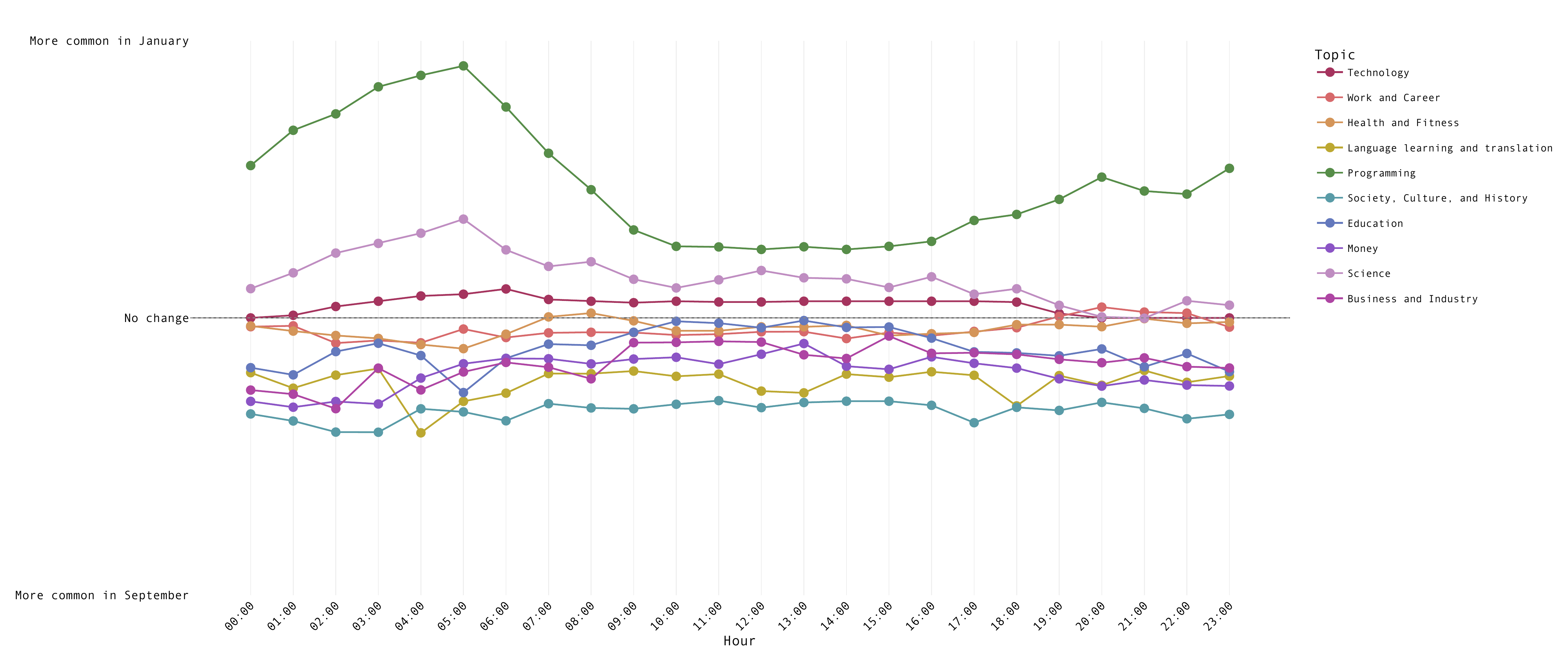}
  \caption{Average hourly rank difference for the top 10 ranks in January vs September, across all devices.}
  \label{fig:jansept_diff}
\end{figure}

In order to further visualize the differences between January and September, we calculate the delta (difference in rank between two months) for the top 10 topics for every hour, averaged across every day (see Figure \ref{fig:jansept_diff}).  

The biggest difference is how ``Programming'' is much more common in January than in September, with the largest variance towards the latter being ``Society, Culture, and History''. This illustrates how, throughout the year, the center of gravity of Copilot usage has shifted from purely productivity-focused conversations towards more social topics. This likely reflects a dual dynamic: the broadening of habits among existing users, and the democratization of the user base as mainstream adopters—who may have less technical priorities than the developer-heavy cohort of early January—joined the platform. This includes, for example, having conversations that range from understanding pop culture to historical events to international politics.

\begin{figure}[p]
    \centering
    \includegraphics[width=\linewidth]{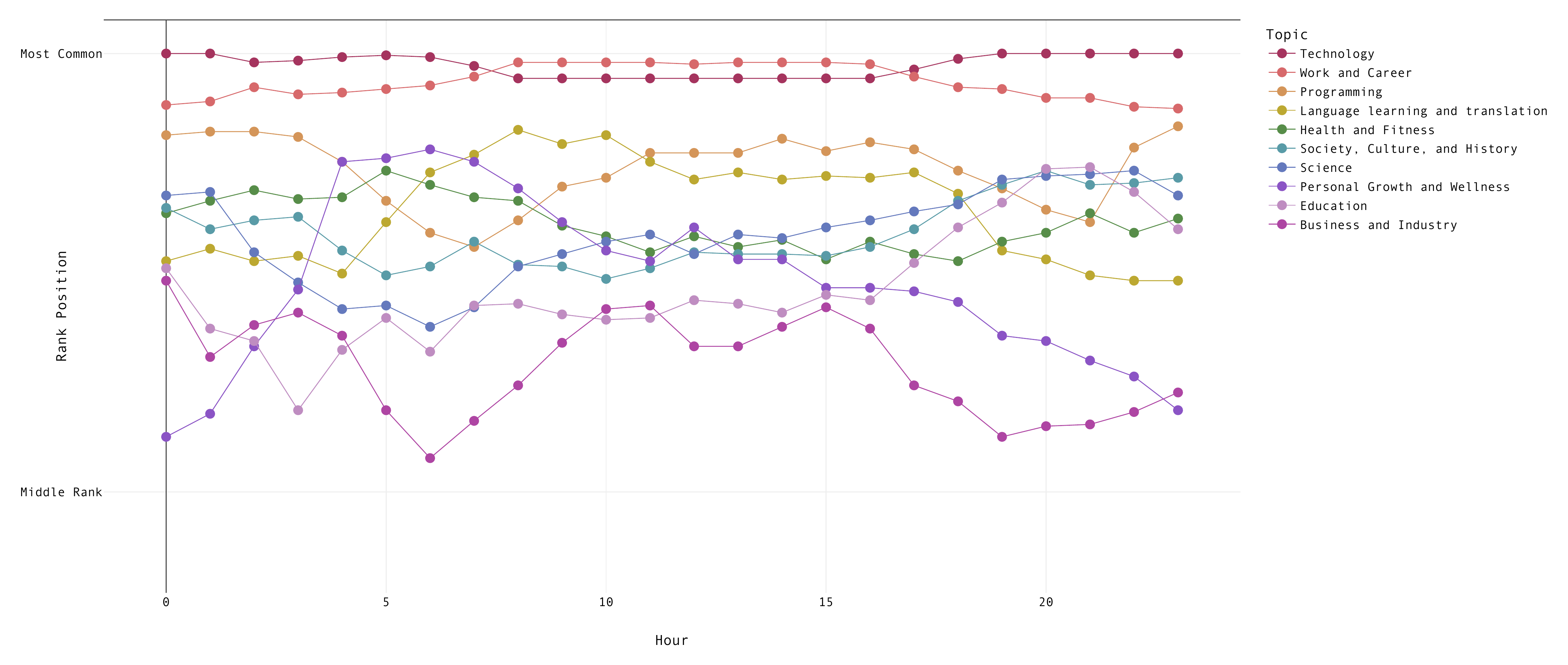}
    \caption{Top 10 average rank of conversation topic per hour, for the month of June, on desktop.}
    \label{fig:june_top_desktop}
\end{figure}

\begin{figure}[htbp]
    \centering
    \includegraphics[width=\linewidth]{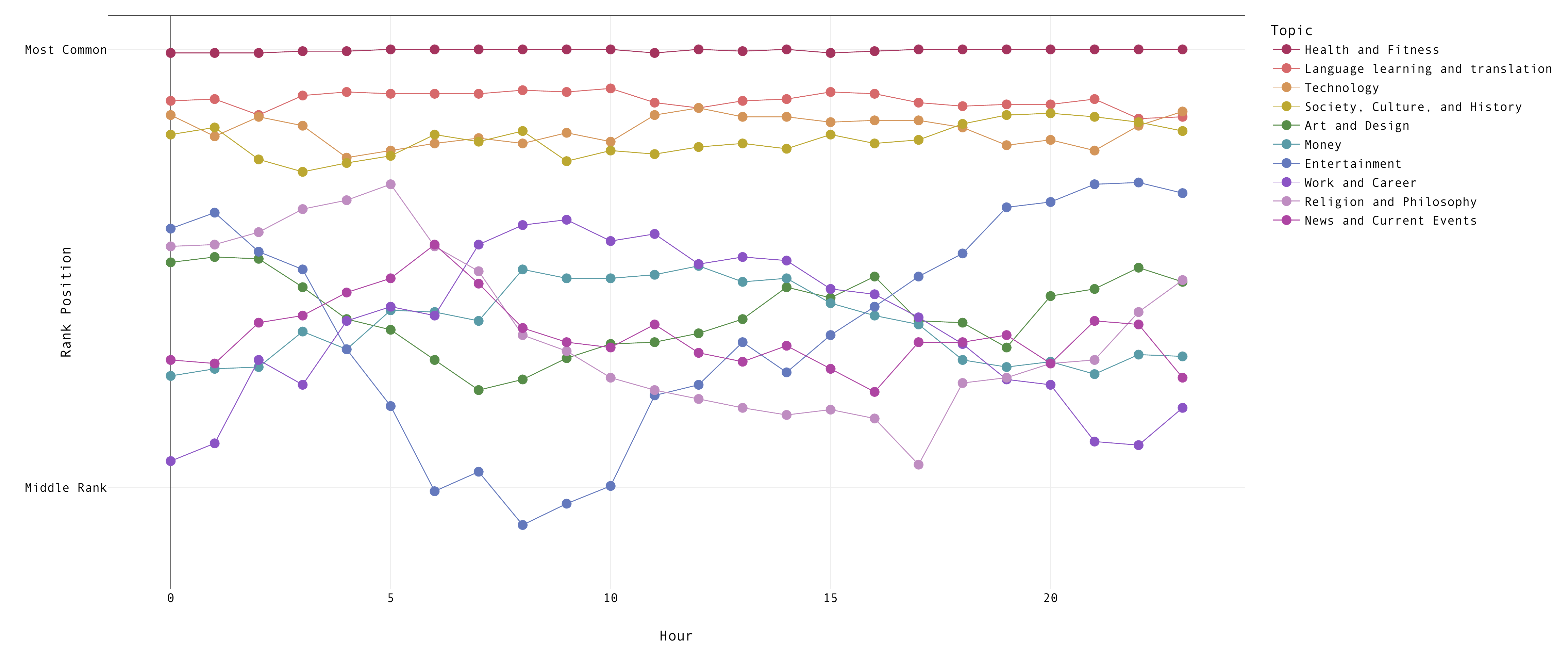}
    \caption{Top 10 average rank of conversation topic per hour, for the month of June, on mobile.}
    \label{fig:june_top_mobile}
\end{figure}

\begin{figure}[htbp]
  \centering
  \includegraphics[width=\linewidth]{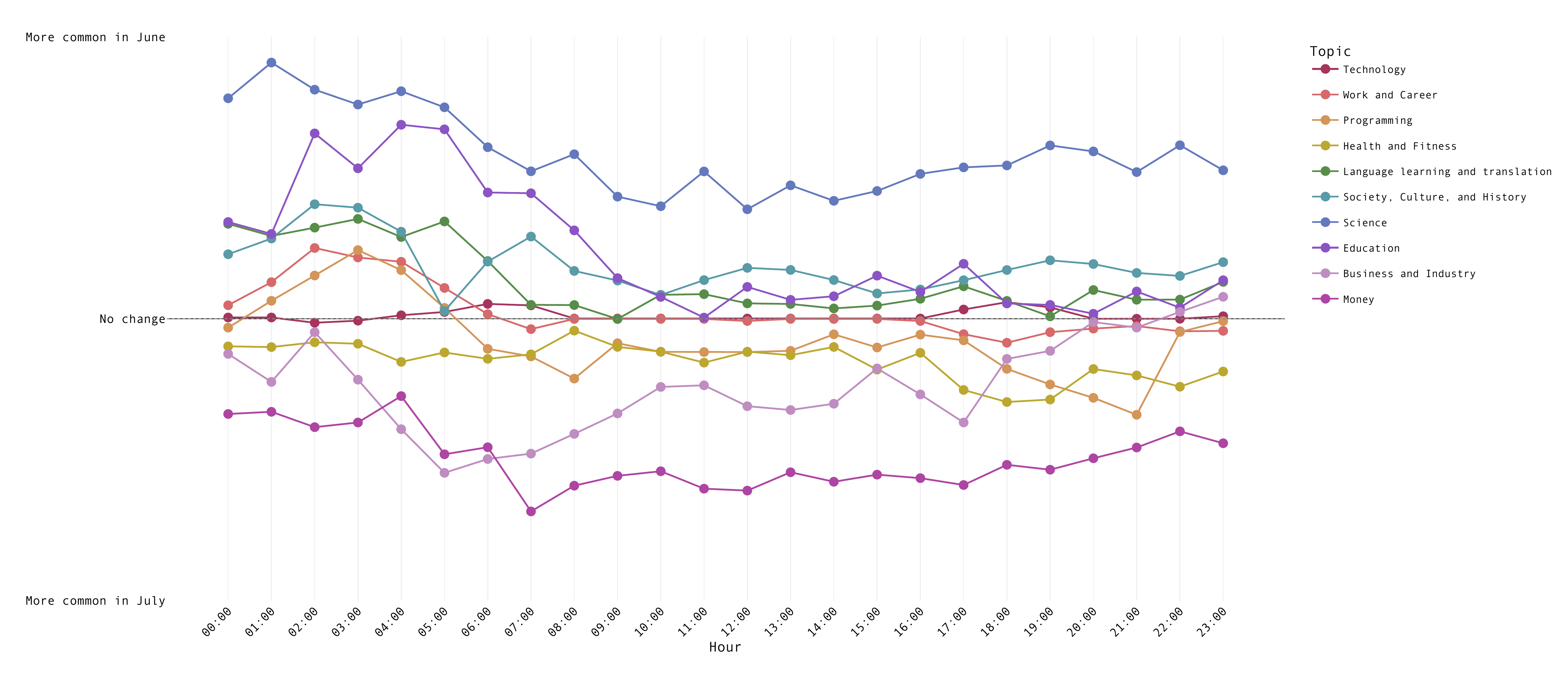}
  \caption{Average hourly rank difference for the Top 10 ranks on desktop, in June vs July.}
  \label{fig:jun_jul_desk}
\end{figure}

\begin{figure}[htbp]
  \centering
  \includegraphics[width=\linewidth]{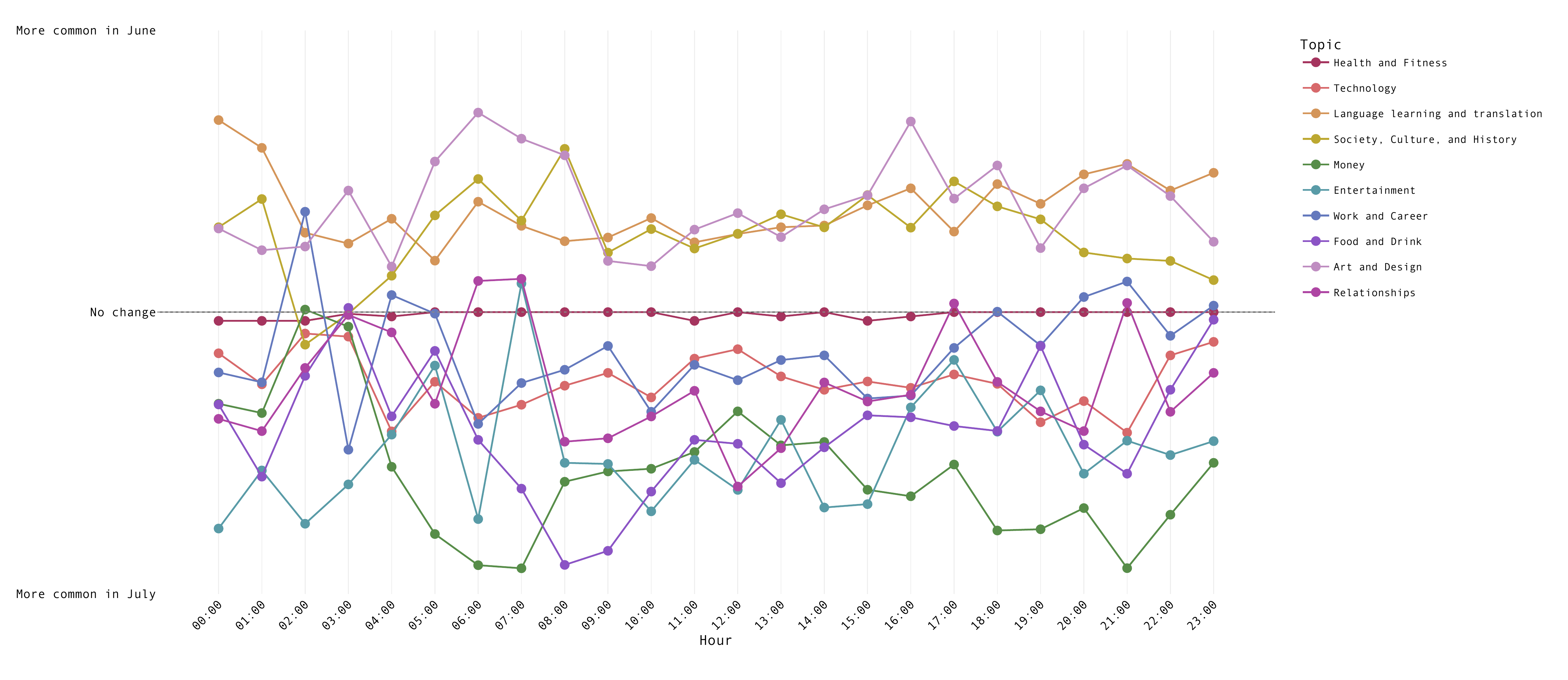}
  \caption{Average hourly rank difference for the Top 10 ranks on mobile, in June vs July.}
  \label{fig:jun_jul_mob}
\end{figure}

It's not just about changes across the year, but also from month to month. For example, two consecutive months that showed a large variation in ranks between them in the previous analysis (in Figures \ref{fig:desktop_tip} and \ref{fig:mobile_tip}) were June and July 2025. For a more in-depth breakdown between these two months, see Figures \ref{fig:june_top_desktop} and \ref{fig:june_top_mobile} for a detailed hourly average of June, and Figures \ref{fig:jun_jul_desk} and \ref{fig:jun_jul_mob} for how that shifted in comparison to July, on desktop and mobile respectively.

By analyzing hourly patterns across devices, we observe three distinct modes of interaction: the workday, the constant personal companion, and the introspective night.

\begin{figure}[htbp]
  \centering
  \includegraphics[width=0.6\linewidth]{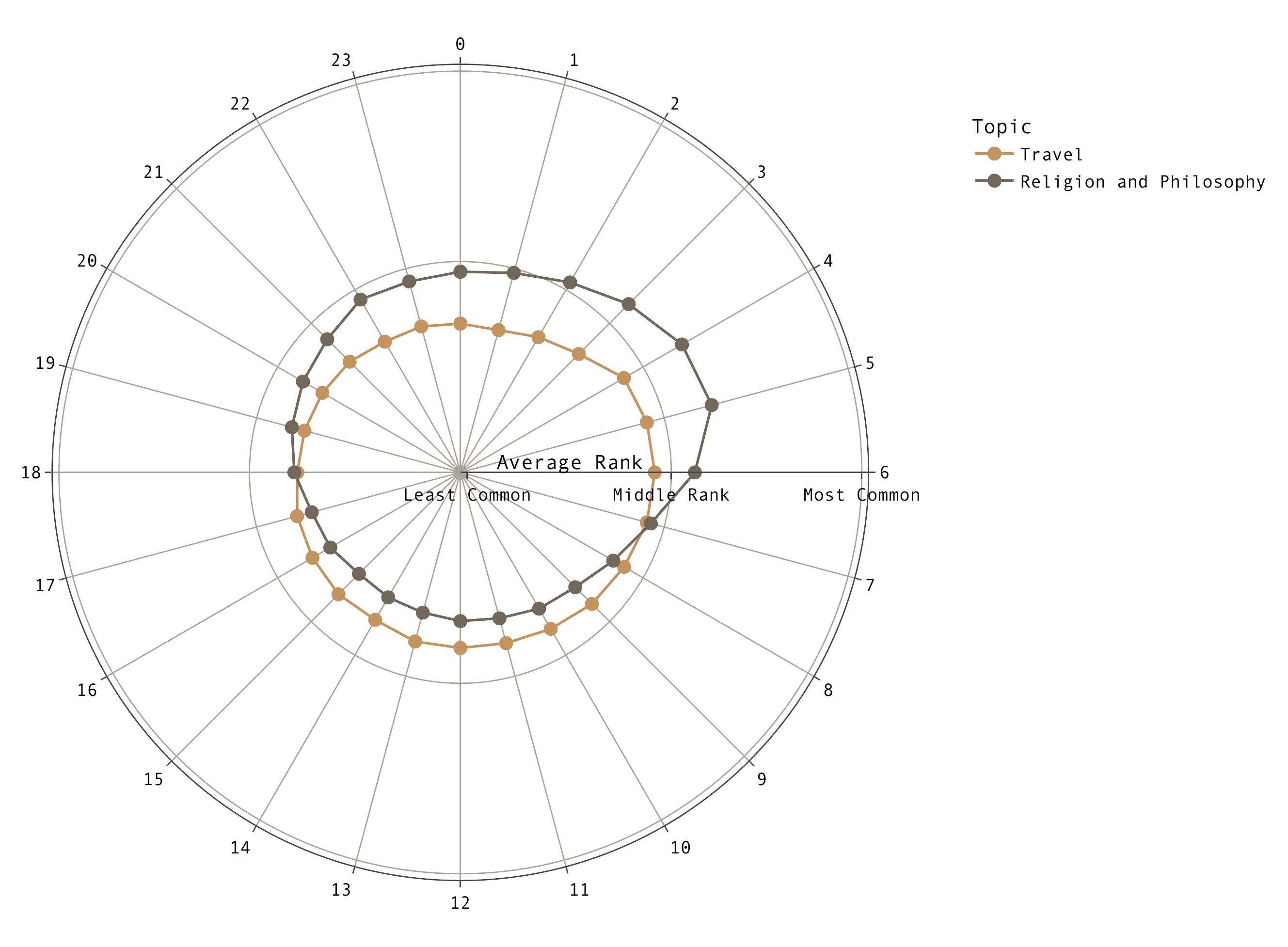}
  \caption{Average hourly rank for Religion and Philosophy and Travel, for all devices, across all months.}
  \label{fig:religiontravel}
\end{figure}

\textbf{The  Workday (8 a.m. to 5 p.m.):}
On desktop, the data mirrors the standard economic workday. Between 8 a.m. and 5 p.m., ``Work and Career'' overtakes ``Technology'' for the top rank, while education and science-focused conversations rise significantly compared to nighttime hours. This professional focus is further highlighted by the inverse pattern on mobile: ``Entertainment'' shows a consistent U-curve, remaining high in the evenings and late nights but dropping substantially during business hours. Interestingly, we also observe a rise in ``Travel'' during these active hours (see Figure \ref{fig:religiontravel}), suggesting that users may be utilizing their work tools to plan trips or prepare for commutes.

\textbf{The Constant Companion:}
While the workday dictates the rhythm for most topics, one subject defies temporal patterns. On mobile, ``Health and Fitness'' remains the most frequent topic across every hour of the day, invariant to the time. This suggests a device-specific usage pattern where the phone serves as a constant confidant for physical well-being, regardless of the user's schedule.

\textbf{The Introspective Night:}
Copilot usage shifts dramatically when the workday ends. As the night deepens, users turn to AI to ask the big questions of life that are not always easy to answer—or perhaps to ask them at a time when others are not awake. ``Religion and Philosophy'' shifts up in rank during the depths of the night through dawn. This late-night introspection is visible across devices, though with monthly nuances; in June, for instance, distinct introspective topics pierced the top 10 that did not appear in July: ``Religion and Philosophy'' on mobile and ``Personal Growth and Wellness'' on desktop.

The idea of looking into daily patterns is not just about the hours but also about the cadence of days. When observing the daily average per month, we found two particularly interesting patterns.

\begin{figure}[htbp]
  \centering
  \includegraphics[width=\linewidth]{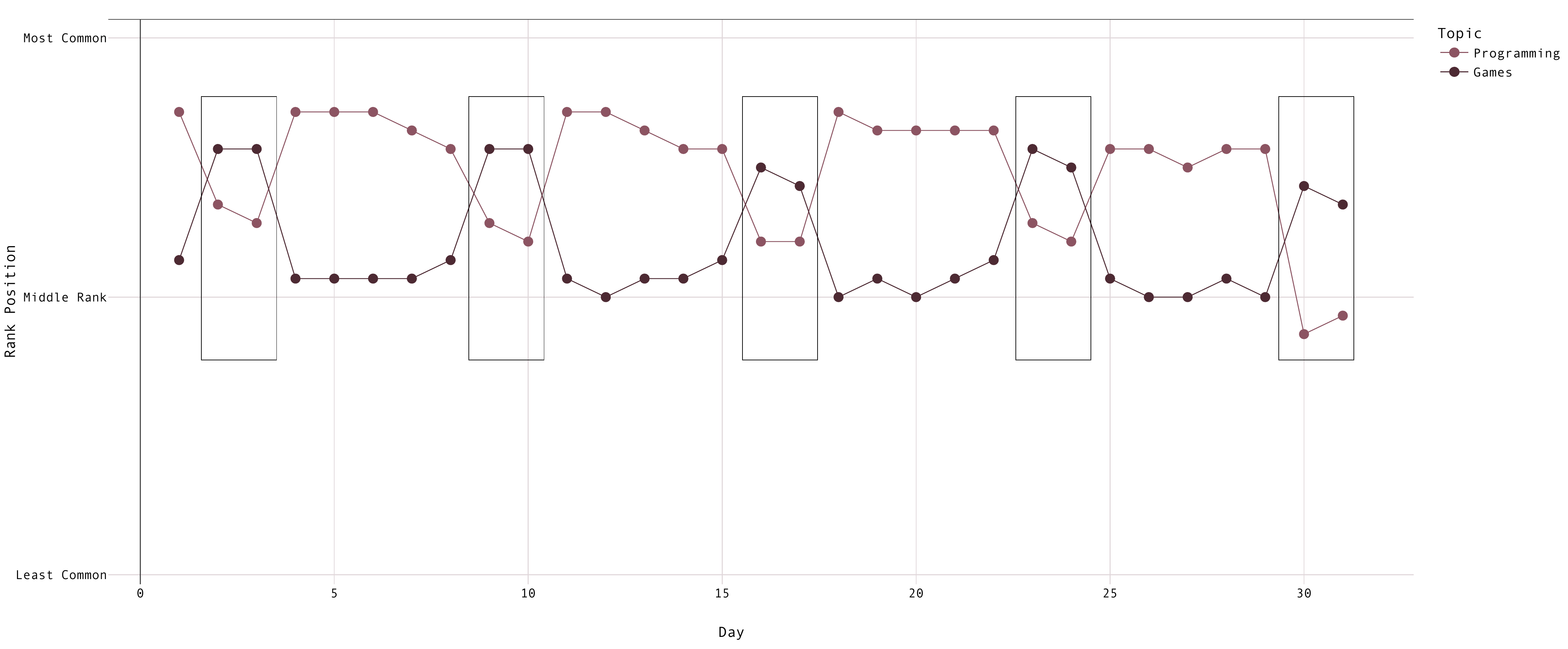}
  \caption{Average daily rank of Programming vs Games conversations, in August. The boxes correspond to the weekends.}
  \label{fig:proggame}
\end{figure}

First, on a direct productivity versus leisure comparison, we found that we can pinpoint which days were weekdays by looking at the comparison between ``Programming'' and ``Games'', as in Figure \ref{fig:proggame}. Further solidifying the usage pattern effects across the day in line with traditional work schedules, ``Programming'' conversations are more common than ``Games'' during the weekdays, and vice versa during the weekend, mirroring each other.

\begin{figure}[htbp]
  \centering
  \includegraphics[width=\linewidth]{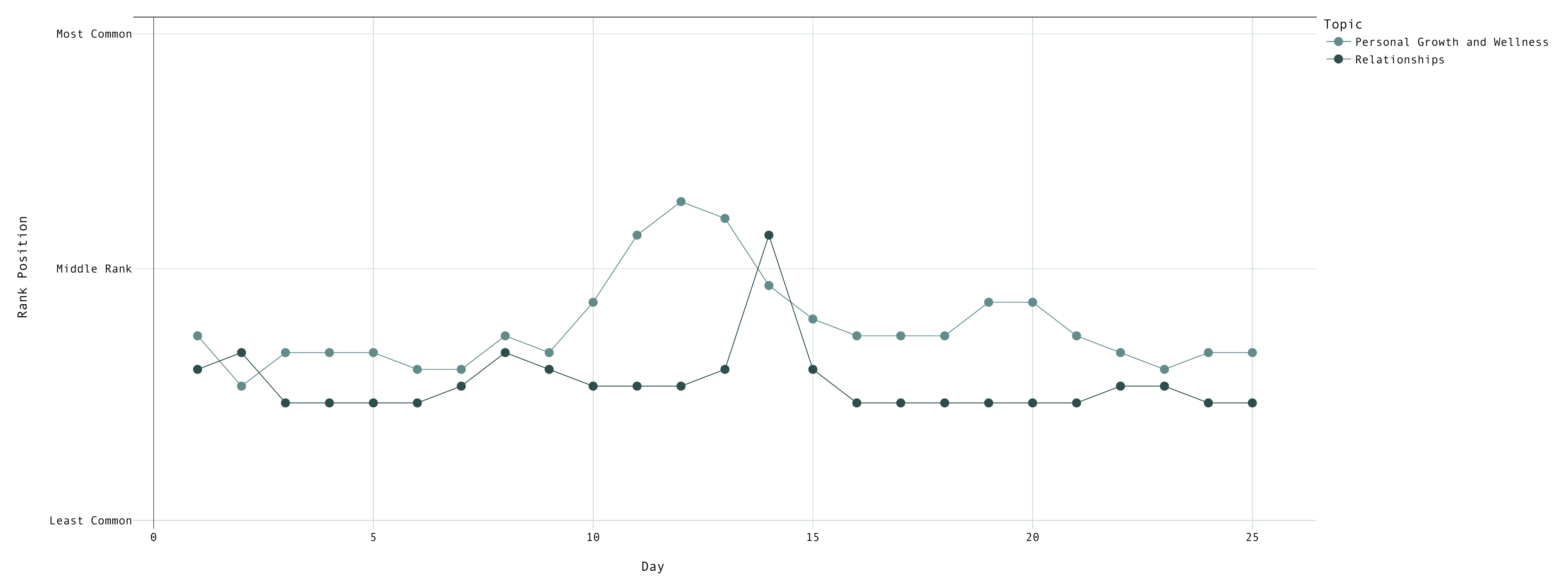}
  \caption{Average daily rank of Relationships and Personal Growth and Wellness, in February.}
  \label{fig:february}
\end{figure}

Second, we see that our users trust Copilot with questions close to their hearts: in the month of February (see Figure \ref{fig:february}), our users had a rise in conversations about ``Personal Growth and Wellness'' in the run up to Valentine's Day, with ``Relationships'' spiking on the day itself.

\section{Discussion}

Our analysis of 37.5 million conversations reveals that Copilot is no longer just a productivity tool; its use is shaped by the temporal and physical context of the user. By disentangling seasonality, daily rhythms, and device-level differences, we move beyond the monolithic view of ``AI usage'' to reveal a technology that has integrated into the full texture of human life. The contrast between the desktop's professional utility and the mobile device's intimate consultation suggests that users are engaging with a single system in two ways: a colleague at their desk and a confidant in their pocket.

This bifurcation has significant implications for the design of generative AI. The industry has largely treated the ``chatbot'' as a uniform experience across endpoints. However, our finding that mobile users prioritize health and fitness—regardless of the hour—indicates that the mobile form factor signals a shift toward personal conversations and self-improvement. This suggests the need for interfaces that are context-aware, differentiating not just in UI, but in personality and capability. A desktop agent should optimize for information density and workflow execution, while a mobile agent might prioritize empathy, brevity, and personal guidance.

Furthermore, the temporal rhythms we observed underscore the generality of Large Language Models, their ability to sync with human circadian rhythms in a way no previous technology has. The rise in philosophical and existential queries during the late night suggests that as the sun sets, the user's need shifts from external productivity (`doing') to internal reflection (`being').

Ultimately, these patterns paint a picture of rapid and deep social integration. Users have tacitly agreed to weave AI into the fabric of their daily existence, turning to it for code reviews at 10 a.m. and existential clarity at 2 a.m. This is not merely a story of adoption, but of adaptation. The data suggests that we are not just using AI to do our work faster; we are using it to navigate the complexities of being human, one prompt at a time.

\FloatBarrier
\printbibliography
\end{document}